\documentclass{jaist}
\usepackage{amsmath,amsfonts}
\usepackage{algorithmic}
\usepackage{algorithm}
\usepackage{array}
\usepackage[caption=false,font=normalsize,labelfont=sf,textfont=sf]{subfig}
\usepackage{textcomp}
\usepackage{stfloats}  
\usepackage{verbatim} 
\usepackage{cite}
\usepackage{amsmath,amssymb}
\usepackage{graphicx}

\usepackage{pdfpages}

\usepackage{xeCJK}

\usepackage{tikz}
\usetikzlibrary{arrows.meta, positioning}
\usepackage{cuted}

\usepackage{listings}
\usepackage{xcolor}  

\usetikzlibrary{positioning}

\usepackage{hyperref} 
 
\volume{4}
\issue{2}
\startpage{1}
\journalname{Journal of Advances in Information Science and Technology}
\publishdate{March 2026}

\shortauthor{L.Huang} 
\shorttitle{ A Calligraphy Score Framework}

\title{Shu Dao: A Calligraphy Score Framework Linking Calligraphy, Music, and Performance}

 \author{
  Lican Huang \textsuperscript{1}
  }
 \jaistthanks{  
    \textsuperscript{1}Hangzhou Domain Zones Technology Co., Ltd., Hangzhou, China. \\
  Corresponding Author: Lican Huang (e-mail:\url{lican.huang.hz@gmail.com}).
  \\
  \\
    Manuscript received March 3, 2026;  revised March 15, 2026;  accepted March 16 , 2026.
}

\begin{document}

\maketitle

 \begin{abstract}

This paper introduces \textbf{Calligraphy Writing Score Representation (CWSR)} and proposes \textit{Shu Dao} as a framework that interprets East Asian calligraphy   as a performative art rather than a static visual artifact. Inspired by traditions such as Japanese \textit{Shodō} and embodied cultural practices such as \textit{Chadao}, the framework models calligraphy as a structured performance analogous to musical notation.

Instead of representing characters as fixed images, the proposed approach encodes each brush stroke as an ordered and executable action, forming a \emph{calligraphy score}. Characters are organized within a structured spatial grid, and strokes are annotated with attributes including stroke type, execution order, spatial coordinates, trajectory, compositional role, and dynamic properties such as brush pressure and pacing. This representation captures temporal and expressive aspects of calligraphic writing that are typically absent from image-based representations.

The paper makes three main contributions. First, it introduces CWSR as a structured notation system for representing calligraphy across multiple levels, including strokes, character structures, and compositional organization (e.g., layout and \textit{zhangfa}), together with their rhythmic and performative dynamics. Second, it conceptualizes \textit{Shu Dao} as a score-mediated framework that models calligraphy as structured performance. Third, it establishes a computational foundation for the analysis, visualization, and executable generation of calligraphic works by AI-based calligraphic agents.

Together, these contributions bridge calligraphy, musical notation, and performative cultural practices, supporting human--AI co-creation in computational calligraphy and digital humanities research.

\end{abstract}  
\jaistkeywords{Shu Dao , Calligraphy Writing Score Representation ,  Calligraphy Score, Music Score, Chadao,  AI-based calligraphic agents }
 \section{Introduction}

East Asian calligraphy is an inherently performative art that integrates brush pressure, speed, direction, rhythm, and full-body movement into a temporally unfolding creative process. Although the resulting characters appear as static visual artifacts, they are fundamentally traces of embodied actions shaped by discipline, intention, and expressive modulation.  Conventional representations—such as printed images or vectorized outlines—collapse the dynamic writing process into static forms. While such representations allow limited image-based modeling and visual reproduction, they omit critical procedural information, including stroke order, rhythm, gesture trajectory, and brush dynamics. As a result, they cannot fully support process-level computational modeling, reproducibility of writing actions, or performative interpretation.

Existing digital approaches to calligraphy have primarily focused on visual reconstruction, stroke extraction, or stylistic imitation. While these methods can reproduce the appearance of written characters, they often fail to encode higher-level structural intent, temporal organization, and performance logic. As a consequence, essential dimensions of calligraphic practice—such as stroke hierarchy, compositional organization, rhythm, and embodied execution—remain implicit or inaccessible within computational frameworks.

To address these limitations, this paper introduces \textbf{Calligraphy Writing Score Representation (CWSR)} and proposes \textbf{\textit{Shu Dao}} as a notation-based conceptual framework for interpreting calligraphy as structured performance. In this work, Shu Dao is defined as a score-based paradigm that views calligraphic writing not merely as visual composition, but as a time-based sequence of expressive actions whose meaning emerges through ordered execution and rhythmic structure.

Within this perspective, characters are treated as performable units analogous to musical measures, and calligraphic writing is interpreted as a score-mediated practice in which brush actions unfold over time according to both structural constraints and expressive interpretation. This conceptualization draws parallels with musical notation as well as embodied cultural practices such as \textit{Chadao}, where structured actions are performed through disciplined yet expressive sequences.

CWSR operationalizes this paradigm through a unified notation system that encodes calligraphy across multiple structural levels. Rather than representing characters as static images, the framework models writing as a structured score composed of ordered brush actions and spatial relationships. Specifically, CWSR captures:

\begin{itemize}
    \item \textbf{Stroke-level representation}, including stroke type, execution order, trajectory, and start and end positions;
    \item \textbf{Character-level structure}, encoding spatial relationships and structural organization;
    \item \textbf{Compositional organization}, including layout and calligraphic composition principles such as \textit{zhangfa};
    \item \textbf{Dynamic and rhythmic attributes}, including brush pressure, pacing, and expressive variation.
\end{itemize}

By transforming calligraphy into a structured and interpretable score, CWSR provides a symbolic intermediary between static inscription and embodied performance. This representation enables computational analysis, visualization, and executable generation of calligraphic works while preserving the temporal and expressive dimensions of writing.

\subsection*{Contributions}

This paper makes three primary contributions:

\begin{enumerate}
    \item It introduces \textbf{Calligraphy Writing Score Representation (CWSR)}, a structured notation system that represents calligraphy across multiple levels, including brush strokes, character structures, compositional organization (e.g., \textit{zhangfa}), and their rhythmic and performative dynamics.

    \item It proposes \textbf{\textit{Shu Dao}} as a score-based conceptual framework that interprets East Asian calligraphy as structured performance rather than static visual form.

    \item It provides a \textbf{computational foundation} enabling analysis, visualization, and executable generation of calligraphic works by AI-based calligraphic agents, supporting new forms of human--AI collaboration in computational calligraphy and digital humanities research.
\end{enumerate}

Through this framework, calligraphy can be understood not only as a visual art but also as a dynamic notation system that bridges artistic tradition, performance practice, and computational creativity.
   
 \section{Related Work and Motivation}

Research on calligraphy, performance, and embodied aesthetics spans multiple disciplinary traditions, including art theory, cultural studies, musicology, and computational modeling. Across these fields, a recurring theme is the recognition that artistic meaning often emerges not solely from static artifacts but from temporally unfolding, embodied action.

From a cultural and philosophical perspective, East Asian practices such as Chinese calligraphy and the tea ceremony are frequently understood as embodied arts in which technique, rhythm, discipline, and moral cultivation are inseparable. Towler characterizes \emph{Cha Dao} as a lived aesthetic system in which form, temporal awareness, and disciplined practice converge into a unified way of life \cite{towler2010cha}. Similarly, Wilson analyzes the Japanese tea ceremony as a pancultural art form, arguing that its artistic value resides not in material artifacts but in ritualized performance and shared embodied meaning \cite{wilson2018japanese}. These perspectives provide an essential conceptual foundation for viewing calligraphy as a process-oriented and performative practice rather than a purely visual product.

Within the broader East Asian calligraphic tradition, writing is often understood not merely as textual inscription but as a cultivated “way” or disciplined practice. In China, this perspective is associated with the philosophical and artistic understanding of calligraphy as a structured, expressive path. In Japan, this notion is formally expressed in the practice of \emph{Shodō} (書道), meaning “the way of writing,” which emphasizes spiritual cultivation, aesthetic discipline, and embodied execution. In both traditions, the act of writing integrates rhythm, brush control, and full-body engagement, producing works that are simultaneously visual and performative.

In this work, the concept of \textit{Shu Dao} is reinterpreted in a computational and representational context. While \emph{Shodō} traditionally emphasizes cultural and spiritual discipline, \textit{Shu Dao} extends this notion by interpreting the “way of writing” as a structured, score-based performance system. This reinterpretation does not replace traditional practice but highlights temporal structure, rhythmic organization, and performative execution, providing a foundation for the proposed Calligraphy Writing Score Representation (CWSR), which encodes calligraphy as an executable sequence of expressive actions.

This interpretation aligns closely with performance-centered theories in musicology. Cook argues that musical meaning is not fully contained within written notation but instead emerges through performance, interpretation, and bodily engagement \cite{cook2013beyond}. Musical scores thus function as mediating structures between composition and execution, specifying order and constraint while allowing expressive variation. Calligraphic works may be interpreted analogously: written characters act as traces of temporally unfolding actions, preserving structural intent while remaining open to interpretative realization. Such insights motivate representational approaches that seek to recover or encode the dynamic processes underlying static inscriptions.

Within calligraphy-specific scholarship, Chiang provides a foundational aesthetic and technical analysis of Chinese calligraphy, emphasizing stroke structure, brush control, balance, and expressive modulation as core evaluative dimensions \cite{chiang1974chinese}. This work remains influential in framing calligraphy as a synthesis of disciplined technique and expressive variation, continuing to inform pedagogical traditions and analytical discourse. However, these principles are typically conveyed descriptively or visually rather than through formal symbolic representations capable of supporting computational modeling or systematic reinterpretation.

Computational modeling of calligraphy and handwriting has developed along several complementary directions. Early systems emphasized stroke-based modeling and simulation, incorporating brush dynamics and haptic feedback to digitally reproduce calligraphic skills \cite{wang2006stroke}. Generative approaches explored the automatic synthesis of artistic Chinese calligraphy using rule-based or learning-based methods, primarily targeting stylistic imitation from exemplar works \cite{xu2005automatic}. In parallel, handwriting recognition research established frameworks for online and offline writing analysis, including stroke segmentation, trajectory modeling, and shape abstraction \cite{824821}. While these studies provide essential technical foundations, their primary objectives often prioritize recognition accuracy, reconstruction fidelity, or visual similarity over interpretability grounded in traditional calligraphic aesthetics and performance logic.

As a result, most existing digital calligraphy systems focus on image-based analysis or vectorized stroke reconstruction. Although effective for reproducing visual appearance, such approaches rarely encode performance intent, temporal modulation, rhythmic organization, or compositional flow. In contrast, musical notation explicitly separates composition from execution while preserving interpretability across performers, contexts, and realizations \cite{cook2013beyond}. A comparable symbolic intermediary has largely been absent in computational treatments of calligraphy.

Motivated by this gap, the present work situates East Asian calligraphy within a broader framework of embodied performance and score-mediated interpretation. By drawing conceptual parallels with music and tea practice \cite{towler2010cha, wilson2018japanese}, and grounding representational design in established calligraphic aesthetics \cite{chiang1974chinese}, this work introduces a score-based representation that operates between static image and embodied act. Rather than replacing existing stroke-based or handwriting models \cite{wang2006stroke, xu2005automatic, 824821}, the proposed framework complements them by elevating interpretability, rhythm, temporal structure, and expressive coherence to first-class analytical dimensions.

This motivation leads to the development of the \textbf{Shu Dao} framework and its operational representation, \textbf{Calligraphy Writing Score Representation (CWSR)}, which reinterpret East Asian calligraphy as a structured, score-based system capable of supporting both human interpretation and computational execution.

\section{Calligraphy Writing Score Representation (CWSR)}

In this framework, \textbf{Shu Dao}, as reinterpreted here, is fundamentally a \textbf{temporal and embodied art}. Every gesture, individual stroke, character, and line of characters unfolds through time, guided by rhythm, bodily control, and expressive intention. The \textbf{Calligraphy Writing Score Representation (CWSR)} models this unfolding process as a structured, time-based score, capturing not only visual form but also execution, rhythm, and artistic performance. Inspired by musical notation and performative practices such as \textit{Chadao}, CWSR treats calligraphy as a multi-layered, executable performance rather than a static image.

 \subsection{Temporal Flow and Three Interwoven Dimensions}

CWSR represents calligraphic writing as a continuous \textbf{temporal flow}
in which three dimensions are encoded simultaneously:

\begin{itemize}
    \item \textbf{Rhythm:} the temporal organization of writing actions,
    including gesture timing, stroke duration, character pacing,
    tempo variation, expressive pauses, and transitions of energy.

    \item \textbf{Calligraphy Actions:} the structured execution of writing
    from preparatory gestures through individual strokes, characters,
    and lines, specifying spatial layout, stroke types, stroke order,
    and structural relations.

    \item \textbf{Artistic Performance:} the expressive and embodied
    qualities of execution, including brush pressure modulation,
    ink density, gesture style, emotional intensity, and whole-body
    engagement.
\end{itemize}

These three dimensions are inseparable: every gesture, stroke, character,
and line is simultaneously rhythmic, structurally organized, and
expressively performed.
 
\subsection{Sequential and Hierarchical Organization in the Calligraphy Score Representation}

CWSR organizes calligraphic writing through a combination of
\textbf{sequential preparatory stages} and a \textbf{hierarchical
structure of writing actions}. The first stages establish the
performative and artistic context of writing, while the subsequent
layers describe the hierarchical structure of calligraphic execution.

\textbf{Sequential preparatory layers}

\begin{enumerate}
    \item \textbf{Preparatory Layer:} establishes the performative state
    prior to writing, including posture, breath control, mental focus,
    and ritualized gestures that align body, mind, and medium.

    \item \textbf{Whole/Art Layer:} defines the global artistic context,
    including paper or silk selection, brush and ink choice,
    composition, margins, and overall aesthetic direction.
\end{enumerate}

\textbf{Hierarchical writing layers}

\begin{enumerate}
    \setcounter{enumi}{2}

    \item \textbf{Line Layer (\textit{Zhangfa}):} organizes characters
    into coherent lines, specifying spatial direction, rhythmic pacing,
    compositional balance, and expressive flow across the work.

    \item \textbf{Character Layer:} treats each character as a structured
    unit composed of multiple strokes, encoding internal balance,
    spatial relationships, stroke order, and structural constraints.

    \item \textbf{Stroke Layer:} represents the elemental writing
    actions, including stroke type, trajectory, start and end points,
    sub-stroke decomposition, and expressive modulation such as
    pressure, speed, and gesture style.
\end{enumerate}

Across these layers, rhythm, structural organization, and artistic
performance remain tightly coupled, enabling CWSR to represent
calligraphy as a unified temporal and performative system that can be
interpreted and executed by both human practitioners and AI agents.

 \subsection{Stroke Relations and Character Integrity}

CWSR explicitly models relational and hierarchical constraints among strokes and characters in order to preserve structural integrity while supporting expressive variation during performance.

\begin{itemize}
    \item \textbf{Adjacency and connectivity:} relationships among stroke start points, end points, midpoints, and directional flow, ensuring coherent transitions between writing actions.
    
    \item \textbf{Layering and overlap:} visual hierarchy and interaction among intersecting strokes within a character, reflecting traditional brush ordering and compositional logic.
    
    \item \textbf{Proportionality and balance:} spatial harmony and compositional stability at both the character and line levels, maintaining visual equilibrium within the evolving structure.
    
    \item \textbf{Character coherence:} constraints ensuring that grouped strokes form well-structured characters, preserving legibility, internal balance, and intended rhythmic flow.
\end{itemize}

Through these relational constraints, CWSR maintains the structural logic of calligraphy while allowing expressive variation in timing, gesture, and pressure. The representation therefore bridges static visual form with the embodied and temporally unfolding performance of writing.
     
 \subsection{\textit{Zhangfa}: Line-Level Performance Structure}

At the line level, \textit{Zhangfa} defines the performative organization of writing,
governing how characters unfold in space and time within a continuous line.

\begin{itemize}
    \item \textbf{Line direction:} vertical or horizontal orientation that shapes both
    visual composition and the gestural trajectory of writing.

    \item \textbf{Rhythmic pacing:} temporal spacing, variation in writing speed, and
    distribution of energy across characters, producing a dynamic rhythm along the line.

    \item \textbf{Compositional coherence:} alignment, proportional balance, and
    expressive tension that connect individual characters into a unified visual and
    performative structure.
\end{itemize}

Through \textit{Zhangfa}, characters are not treated as isolated units but as elements
within a continuous temporal and expressive sequence. The line thus becomes a higher-level
performance structure in which rhythm, spatial composition, and embodied gesture interact
to shape the overall flow of the calligraphic work.

\subsection{Performance Analogy with Music and \textit{Chadao}}

CWSR situates calligraphy within a broader landscape of performative arts, emphasizing
its nature as an embodied and executable practice rather than a purely static visual
artifact. From this perspective, writing can be understood as a temporal performance
whose structure emerges through execution.

Within this analogy:

\begin{itemize}
    \item Stroke sequences correspond to musical notes as discrete events unfolding in time.
    \item Variations in brush pressure and movement speed function analogously to musical
    dynamics and tempo.
    \item Characters form structured groupings comparable to musical phrasing or measures.
    \item Preparatory gestures, posture, and breath parallel the ritualized bodily engagement
    central to \textit{Chadao}.
\end{itemize}

This analogy highlights how calligraphy, like music and tea practice, integrates
structure, rhythm, and embodied action within a unified artistic performance.
   
 \subsection{Summary of the CWSR Framework}

The Calligraphy Writing Score Representation (CWSR) provides a unified framework for
modeling calligraphy as a structured and temporally unfolding process. Rather than
representing calligraphic works solely as static images, CWSR captures the dynamic
logic of writing through a multi-layered score that encodes rhythm, structural
composition, and embodied execution.

The framework organizes calligraphic practice through sequential preparatory stages and hierarchical writing layers, ranging from preparatory actions and global artistic configuration to line organization (\textit{Zhangfa}), character structure, and elemental strokes. At each level, CWSR integrates three inseparable dimensions: temporal rhythm, calligraphic structure, and artistic performance. These dimensions describe how writing unfolds through time, how characters are structurally formed, and how expressive qualities—such as pressure, speed, and gesture—shape the resulting work.

By explicitly modeling stroke relations, character integrity, and line-level
composition, CWSR preserves the structural logic of traditional calligraphy while
allowing expressive variation in execution. The resulting representation functions
analogously to a musical score, providing a symbolic intermediary between artistic
conception and performative realization.

Through this score-based representation, CWSR reinterprets \textbf{Shu Dao}—the
“Way of Writing”—as a computationally representable and executable system. This
perspective bridges traditional calligraphic aesthetics with modern computational
methods, enabling reproducible analysis, digital preservation, and AI-assisted
interpretation of calligraphic performance.

 \section{From Static Calligraphy to Performable Score}

Traditional Chinese calligraphy is often preserved through completed works,
rubbings, and model books. While visually precise, such representations
obscure essential aspects of the writing process, including stroke timing,
rhythmic pacing, pressure modulation, and the embodied gestures of the
calligrapher. As a result, calligraphy is frequently perceived as a static
visual artifact, even though its creation unfolds dynamically through time,
motion, and expressive intention.

In the philosophy of \textit{Shu Dao} (the “Way of Writing”), calligraphy is
understood as a performative art in which rhythm, bodily movement, and
expressive energy play a central role. From this perspective, the written
form can be viewed as the visible trace of a temporally structured sequence
of actions.

Building on this idea, the CWSR framework reconceptualizes calligraphy as a
\emph{performable score}. Similar to musical notation or ritual sequences in
\textit{Chadao}, the score encodes ordered actions, spatial structure, and
expressive parameters while allowing multiple interpretations. This
score-based representation enables both human artists and computational
systems to interpret and perform the same underlying calligraphic structure.
             
\subsection{From Score Template to Human and AI Performance}

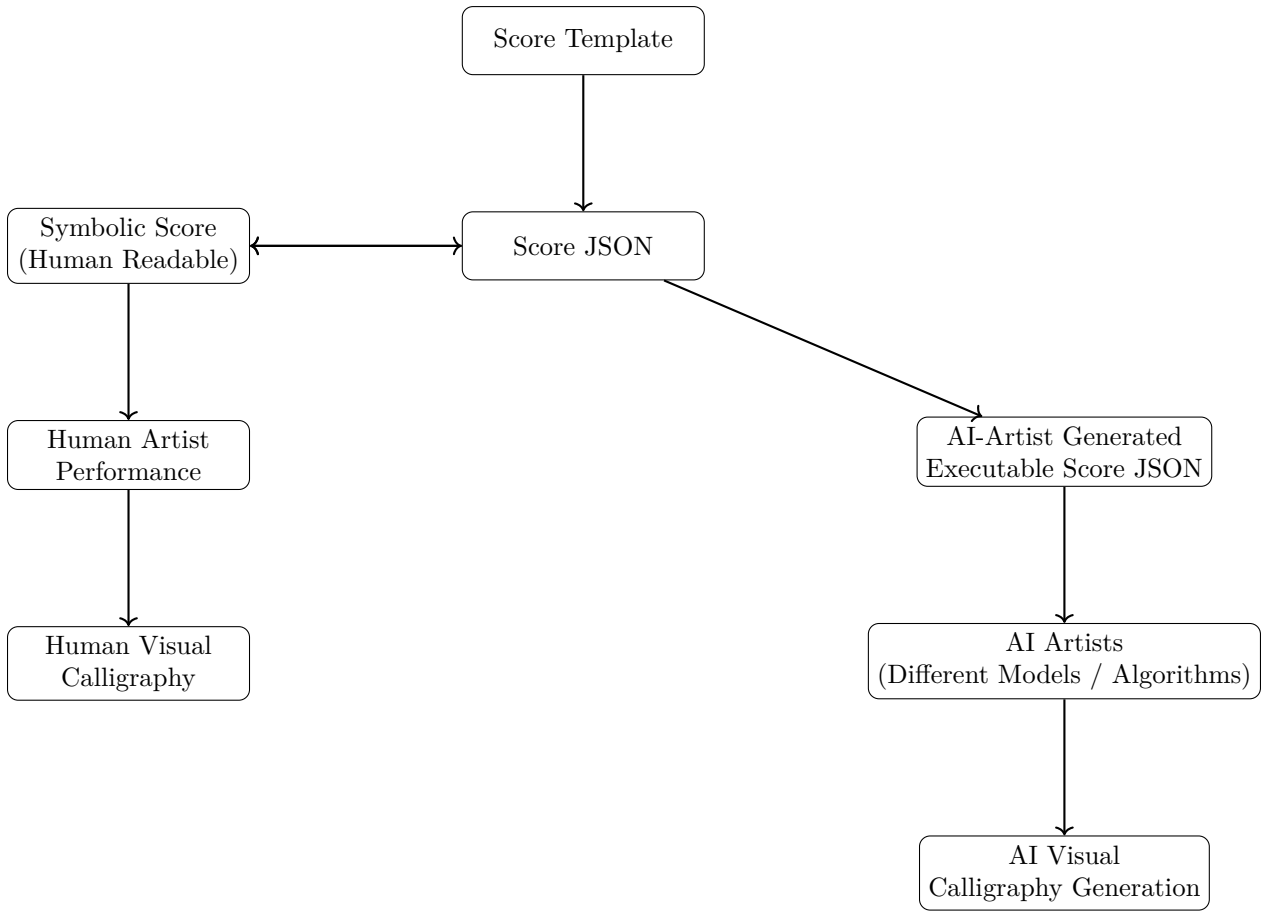
\begin{figure*}[t]
\centering
\begin{tikzpicture}[
node distance=1.8cm and 2.8cm,
box/.style={draw, rectangle, rounded corners, align=center, minimum width=3.2cm, minimum height=0.9cm},
arrow/.style={->, thick}
]

\node[box] (template) {Score Template};

\node[box, below=of template] (json) {Score JSON};

\node[box, left=of json] (symbolic) {Symbolic Score \\ (Human Readable)};

\draw[<->, thick] (json) -- (symbolic);

\node[box, below right=of json] (aijson) {AI-Artist Generated \\ Executable Score JSON};

\node[box, below=of symbolic] (humanperf) {Human Artist \\ Performance};

\node[box, below=of aijson] (aigen) {AI Artists \\ (Different Models / Algorithms)};

\node[box, below=of humanperf] (humanvis) {Human Visual \\ Calligraphy};

\node[box, below=of aigen] (aivis) {AI Visual \\ Calligraphy Generation};

\draw[arrow] (template) -- (json);

\draw[arrow] (json) -- (symbolic);
\draw[arrow] (json) -- (aijson);

\draw[arrow] (symbolic) -- (humanperf);
\draw[arrow] (aijson) -- (aigen);

\draw[arrow] (humanperf) -- (humanvis);
\draw[arrow] (aigen) -- (aivis);

\end{tikzpicture}

\caption{CWSR calligraphy score pipeline. A score template is instantiated as a machine-readable Score JSON. From this representation, two parallel interpretation paths emerge. Human artists interpret the symbolic score to produce diverse calligraphic performances, while different AI artists (models or algorithms) generate executable score variants that lead to computational calligraphy generation.}
\label{fig:cwsr_pipeline}
\end{figure*}

Within the CWSR framework, the calligraphy score functions as a symbolic intermediary
between conceptual design, executable representation, and artistic performance.
 The overall process can be understood as a multi-stage transformation pipeline,
as illustrated in Fig.~\ref{fig:cwsr_pipeline}, which shows how a calligraphy
score is instantiated, interpreted by human and AI artists, and ultimately
realized as diverse visual calligraphic outcomes.
First, a \textbf{Score Template} defines the structural schema of the calligraphy score,
including hierarchical layers, stroke attributes, rhythmic parameters, and relational
constraints. This template establishes the formal structure for representing the writing
process.

Second, the template is instantiated as a machine-readable \textbf{Score JSON}, which
encodes the ordered sequence of calligraphic actions, temporal rhythm, spatial relations,
and expressive parameters. The Score JSON serves as the central representation of the
CWSR framework.

From this representation, two parallel interpretation paths emerge.

\textbf{Human interpretation.}  
 The Score JSON can be rendered as a  \textbf{symbolic calligraphy score}  for human artists, presenting the structural information in a human-readable format analogous to musical notation. The symbolic score and the Score JSON encode the \emph{same underlying structure} but differ only in representation. Using this score, human calligraphers can interpret and perform the work with individual expressive variation while preserving structural and rhythmic coherence.

Different human artists may produce distinct calligraphic realizations from the same
score. Variations in skill level, training traditions, stylistic preferences, and
embodied brush control naturally lead to diverse visual outcomes, even when the
underlying structural instructions remain identical.

\textbf{AI interpretation.}  
In parallel, the same Score JSON can be processed by computational systems to generate
multiple \textbf{AI-Artist-Generated Executable Score JSON} variants. In this context,
an \textbf{AI artist} refers to a computational model or generative system capable of
interpreting and transforming the calligraphy score according to specific stylistic
or algorithmic principles.

Different AI artists may employ diverse generative algorithms, learned style models,
or parameter configurations to produce alternative executable scores. These generated
scores may introduce stylistic, temporal, or expressive variations while maintaining
the structural constraints defined by the   original Score JSON.

Both human-performed and AI-generated score realizations can subsequently be rendered
into \textbf{visual calligraphy generation}, producing distinct visual or animated
representations corresponding to different interpretations of the same underlying
score.

Consequently, a single CWSR score may give rise to multiple calligraphic realizations,
produced either by different human performers or by different AI artists. In this sense,
the score functions similarly to musical notation: it defines the structural and temporal
logic of the work, while the final visual form emerges through interpretation and
performance.

From this perspective, CWSR may be interpreted as a computational formulation
of \textbf{Shu Dao}—the “Way of Writing”—where the essence of calligraphy is
captured not as a static image but as a structured sequence of temporally
organized actions that can be interpreted and performed by both human and
AI artists.

 \section{Example Experiment: A Simple Demo of CWSR}

To demonstrate the practical implications of CWSR, we conducted an exploratory experiment in which a calligraphy score was used to generate visual output computationally. This experiment highlights how a temporal, performable representation can fully encode expressive, embodied brushwork.

\subsection{From Score Template to Score JSON}

 A CWSR score for the canonical characters ``永和九年'' was manually specified, capturing grid-based stroke positions, execution order, and adjacency relationships. The spatial structure of each character is defined within a $9\times9$ reference grid, which provides normalized coordinates for stroke placement and structural constraints.

No reference images were used; the final visual forms emerge entirely from the score description encoded in the CWSR representation. An example for the character ``永'' is shown below, illustrating how strokes are specified using grid positions, ordered execution, and relational constraints.

\begin{strip}
 
\begin{verbatim}
 
# -----------------------------
# Example template
# -----------------------------

# -----------------------------
# Global metadata (SCORE_METADATA)
# -----------------------------
SCORE_METADATA = {
    "descriptive": {
        "Beginning": {
            "title": "Shu Dao Score",
            "style": "Wang Xizhi"
        }
    },
    "Art/Whole": {
        "paper": {
            "type": "silk imitation",
            "texture": "fine fiber",
            "absorbency": "medium-low"
            # Optional: margin can be auto-filled by generator
        },
        "layout": {
            "direction": "top-to-bottom",
            "line_direction": "right-to-left"
            # Optional: line_length, paragraph_gap auto-filled
        },
        "brush": {
            "type": "traditional_chinese_brush",
            "size": 6.0
        },
        "ink": {
            "type": "carbon_ink"
        },
        "rhythm": {
            "time_basis": "continuous",
            "self_rhythm_time": 100
        },
        "performance": {
            "gesture_style": "controlled_fluid",
            "motion_origin": "wrist_elbow",
            "force_profile": "variable_controlled"
        },
        "pre_performance_expression": {
            "vocalization": {
                "type": "chanting_or_exclamation",
                "intensity": "low_to_moderate"
            },
            "body_engagement": "whole_body",
            "breath_state": "deep_regulated"
        },
        "performance_effect": {
            "energy_flow": "coherent",
            "expressiveness": "restrained_intense",
            "momentum": "internally_driven"
        }
    }
}

# -----------------------------
# Example strokes per character
# -----------------------------
example_strokes_characters = {
    
    "永": {
        "strokes": [
            { "type": "dian", "start": [3, 5], "end": [3, 5] },
            {
                "type": "heng-zhe-gou",
                "segments": [
                    { "subtype": "heng", "start": [4, 4], "end": [4, 5] },
                    { "subtype": "zhe",  "start": [4, 5], "end": [7, 5] },
                    { "subtype": "gou",  "start": [7, 5], "end": [6, 4] }
                ]
            },
            {
                "type": "heng-pie",
                "segments": [
                    { "subtype": "heng", "start": [5, 3], "end": [5, 4] },
                    { "subtype": "pie",  "start": [5, 5], "end": [7, 4] }
                ]
            },
            { "type": "pie", "start": [4, 6], "end": [5, 5] },
            { "type": "na",  "start": [5, 5], "end": [7, 7] }
        ],
        "Structure": {
            "comment": "中宫紧收，欹正相生",
            "character_size": 1.0,
            "connections": {
                "end_start": [[4, 5]],
                "middle_end":[[2,4]],
                "middle_start":[[2,5]]

            },
         
            "hint": {
                "motion": "flowing",
                "pressure": "crescendo",
                "gesture": "press-lift-turn",
                "emphasis": "central"
            }
        }
    },
     "和": { ... },
    ...
}
\end{verbatim}
\end{strip}

The annotated strokes were then compiled into a hierarchical JSON-based CWSR score file (see Appendix~A.1).
 
  \subsection{From Score JSON to Symbolic Score}

The JSON-based CWSR scores can be rendered as a \emph{human-readable symbolic score},
analogous to musical notation. While the Score JSON provides a machine-readable
representation of the calligraphy score, the symbolic score presents the same
underlying structure in a visual format suitable for human interpretation.

This representation makes temporal, rhythmic, and performative aspects of
calligraphy explicit, visualizing:

\begin{itemize}
    \item \textbf{Stroke sequence:} the ordered execution of strokes within each character.
    \item \textbf{Temporal rhythm:} inter-stroke intervals, phrasing, and suggested pauses.
    \item \textbf{Gestural cues:} mapping strokes to wrist, arm, or whole-body motion.
    \item \textbf{Expressive intent:} pressure variation, ink modulation, stylistic emphasis, and emotional tone.
    \item \textbf{Line-level organization:} \textit{Zhangfa}, guiding the spatial and temporal arrangement of characters.
\end{itemize}

The symbolic score therefore acts as a bridge between abstract score
representation and performable calligraphic practice. It allows the same
underlying score to be interpreted by human calligraphers or computational
systems, enabling multiple realizations of the work (see Appendix~A.2).
   
  \subsection{From Score JSON to AI-Artist-Generated Executable Score JSON}

Beyond manual authoring, CWSR scores can also be generated by an
\textbf{AI artist} operating directly at the symbolic score level.
In this setting, the AI artist does not synthesize pixels or vector
outlines; instead, it produces an \emph{executable CWSR score}—a
structured sequence of performable calligraphic actions that can be
rendered, interpreted, or executed by downstream systems.

Conceptually, the AI artist functions analogously to a human
calligrapher. It selects stroke features, determines execution order,
assigns spatial relations within the grid, and modulates dynamic
attributes such as rhythm, pressure, and pacing. The resulting
output is a \textbf{complete executable Score JSON} that preserves
character identity while allowing expressive variation, making the
generation process both interpretable and controllable.

Importantly, \textit{Shu Dao} supports \textbf{multiple AI artists}
rather than a single generative model. Each AI artist may embody
distinct stylistic tendencies, learning histories, and evaluation
thresholds, analogous to different calligraphic schools or masters.
These thresholds define acceptable ranges for balance, stroke
tension, rhythmic regularity, and expressive deviation, enabling
diverse yet structurally valid score realizations.

Unlike human artists, AI artists provide \textbf{reproducibility}.
Given the same initial conditions, constraints, and random seed,
an AI artist will produce the same executable score. This property
enables systematic comparison, analysis, and iterative refinement,
with variation introduced through explicit parameter adjustments
rather than uncontrolled visual randomness.

Furthermore, AI artists may \textbf{optimize directly at the score
level}. Through repeated generation and evaluation, an agent can
refine stroke sequencing, temporal pacing, and expressive
modulation, improving execution quality while remaining grounded
in the symbolic structure of CWSR. Learning thus occurs in terms
of \emph{how strokes are performed}, rather than merely how final
images appear.
 
By enabling AI agents to generate executable calligraphy scores, CWSR provides a principled foundation for interpretable and reproducible AI artists capable of stylistic development. Because the representation encodes calligraphy as structured performable actions rather than static images, it naturally bridges human calligraphic practice, computational modeling, and performative generation (see Appendix~A.3). 
 
 \subsection{From AI-Artist-Generated Executable Score JSON to Visual Calligraphy Generation}
 Given an executable CWSR score, a visual calligraphy rendering can be
produced by reconstructing the spatial form of the characters from the
structured instructions encoded in the score.

In the current implementation, rendering relies primarily on the
grid-based spatial coordinates specified in the executable Score JSON.
Other features automatically filled by the AI agent—such as motion hints,
pressure curves, and expressive gestures—are only partially utilized or
currently omitted.
 
\begin{itemize}
    \item \textbf{Stroke reconstruction:} strokes are plotted according
    to grid coordinates, with complex strokes decomposed into ordered
    sub-strokes.

    \item \textbf{Stroke ordering:} strokes follow the execution sequence
    specified in the executable Score JSON, preserving structural integrity.

    \item \textbf{Spatial composition:} characters are positioned according
    to the grid layout and line-level arrangement.

    \item \textbf{Line-level coherence:} characters are arranged according
    to \textit{Zhangfa}, maintaining compositional balance and flow.
\end{itemize}

The Appendix A.4   shows  Python code that  demonstrates how characters can be
rendered in SVG format directly from an executable Score JSON.
It interprets stroke order, spatial layout, and optional stylistic hints
produced by AI agents. While the current implementation primarily uses
grid-based spatial information, the code structure supports the inclusion
of additional performative attributes such as rhythm, pressure, and gesture dynamics.

\begin{figure*}[t]
\centering
\includegraphics[width=0.9\textwidth]{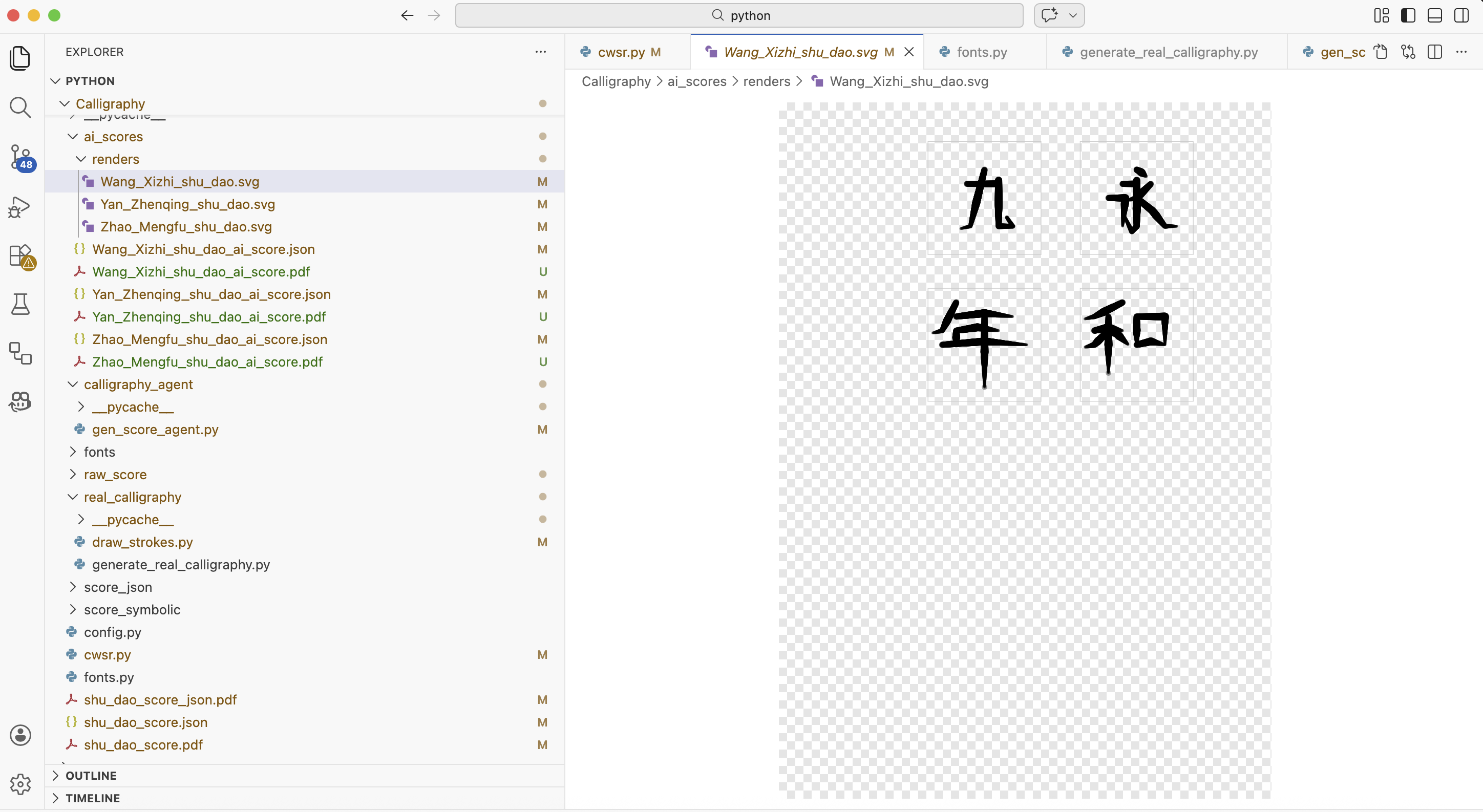}
\caption{Example calligraphy rendering of ``永和'' and ``九年'' generated
from one AI agent’s executable Score JSON . The figure also shows the
experiment directory on the left, containing the three executable Score JSON
files and the three corresponding calligraphy renderings produced by independent AI agents , demonstrating multiple stylistic outputs
from the same Score JSON.}
\label{fig:shu_dao_yong}
\end{figure*}

Figure~\ref{fig:shu_dao_yong} illustrates generated calligraphy for the
characters ``永和'' and ``九年'', produced entirely from an executable
Score JSON without the use of reference images. 

Three independent AI agents were used to generate executable CWSR scores
from the same initial Score JSON. Each agent produced a distinct
\textbf{executable Score JSON}, encoding its own interpretation of the
calligraphic actions. All three outputs are stored in the experiment
directory, which is displayed on the left side of Figure~\ref{fig:shu_dao_yong}. 
On the right, a representative rendering illustrates one of the generated calligraphies.

This demonstration highlights three key capabilities of CWSR:

\begin{itemize}
    \item \textbf{Reproducibility:} each executable Score JSON produces a
    consistent visual outcome.

    \item \textbf{Interpretability:} the executable Score JSON explicitly
    encodes stroke order and spatial layout, making the generative process
    transparent.

    \item \textbf{Generativity:} multiple AI agents can produce distinct
    executable Score JSON files for the same text, supporting stylistic
    variation.
\end{itemize}

Even though the current rendering focuses on spatial reconstruction,
the framework is designed to incorporate additional performative
attributes such as rhythm, pressure, and gesture dynamics in future
visualizations . 
    
\section{Discussion}

This work frames calligraphy not merely as a visual artifact but as a
time-based and embodied practice. By introducing Calligraphy Writing Score
Representation (CWSR), we propose a new perspective for analyzing,
preserving, and computationally modeling calligraphy. This section discusses
key design choices, interpretive implications, practical applications, and
future directions.

\subsection{Stroke Type Taxonomy}

CWSR builds on canonical calligraphic strokes, including simple forms
(\textit{dian}, \textit{heng}, \textit{shu}, \textit{pie}, \textit{na}) and
compound strokes (e.g., \textit{heng--zhe--gou}). Stroke types function as
structural primitives analogous to musical note classes: they define the
skeleton of a character while allowing expressive variation in execution,
including speed, pressure, curvature, and stylistic emphasis.
 
 \subsection{Grid Resolution and Spatial Design}

Traditional calligraphy pedagogy often employs a $3\times3$ grid to guide
character balance, organizing strokes around a central axis with roughly
balanced spatial quadrants. In CWSR, this idea is extended into a
configurable grid-based coordinate system. For example, a $9\times9$ grid
may be used to provide finer spatial resolution while preserving the
structural principles of traditional practice.

Rather than imposing rigid geometry, the grid functions as a symbolic
coordinate system that supports spatial reasoning while remaining
style-agnostic. The grid resolution is configurable: lower-resolution
grids provide greater interpretive freedom, while higher-resolution
grids enable more precise spatial encoding of stroke placement.

\subsection{Implementation Considerations}

CWSR scores may be authored manually, captured via digitizing tablets,
or inferred from handwriting datasets. The JSON-like serialization used
in this work prioritizes readability and interoperability, but the
representation itself is not tied to a specific format; graph-based,
tensor-based, or other structured encodings can be used within the same
conceptual framework.

\subsection{Interpretive Depth}

Unlike conventional calligraphy datasets that store only static images,
CWSR captures multiple layers of performative information:

\begin{itemize}
    \item temporal rhythm and pacing across strokes and characters,
    \item embodied gestures associated with wrist, arm, or whole-body movement,
    \item material affordances such as brush type, paper absorbency, and ink behavior,
    \item expressive states during execution, including hesitation,
    acceleration, and pressure modulation.
\end{itemize}

As with a musical score, the representation does not prescribe a single
visual outcome. Meaning emerges through execution and interpretation rather
than from a static form alone.

\subsection{Implications for Computer Vision and Machine Learning}

CWSR provides a structured intermediate representation between raw
visual data and high-level semantic interpretation. By explicitly encoding
stroke order, temporal dependencies, and performative dynamics, it enables:

\begin{enumerate}
    \item supervisory signals for generative models that encourage
    style-consistent synthesis,
    \item compact representations for style comparison, clustering,
    or authorship analysis,
    \item evaluation metrics that assess performative correctness
    beyond simple visual similarity.
\end{enumerate}

These capabilities allow learning systems to reason not only about
\emph{what} a character looks like, but also \emph{how} it is executed.

\subsection{Performance and Embodied Knowledge}

A central contribution of CWSR is the reframing of calligraphy as a
performative art. Similar to music performance or \textit{Chadao},
calligraphic mastery involves timing, rhythm, and bodily awareness in
addition to visual precision. By encoding these dimensions explicitly,
CWSR enables the transmission and computational analysis of embodied
knowledge that is otherwise difficult to capture.
 
 \subsection{AI Artists and Executable Calligraphy Scores}

Within the Shu Dao framework, AI systems operate not merely as image
generators but as \textbf{AI artists} that produce executable calligraphy
scores. Instead of synthesizing pixels directly, an AI artist generates a
executable  Score JSON describing how a piece of calligraphy should be performed.
The final visual result emerges only through interpretation and execution
of the score, mirroring the role of a human calligrapher.

This distinction has important implications. Whereas most computational
approaches treat calligraphy as a problem of visual synthesis or style
transfer, AI artists in CWSR engage with calligraphy as a \textbf{temporal
and embodied process}. Stroke order, rhythm, pacing, and expressive
modulation are explicitly represented in the generated score, making the
generative process interpretable and inspectable.

The framework naturally supports \textbf{multiple AI artists}. Different
agents may embody distinct stylistic tendencies, learned constraints,
or evaluation criteria, analogous to different schools or masters in
traditional calligraphy. Consequently, the same textual content can
produce multiple structurally valid but stylistically distinct score
realizations.

Unlike human practitioners, AI artists also provide
\textbf{reproducibility}. Given identical initial conditions and random
seeds, an AI artist produces the same executable score. This property
enables controlled experimentation, systematic comparison of stylistic
strategies, and reproducible evaluation.

Finally, AI artists may evolve through \textbf{score-level optimization}.
Rather than learning solely from visual similarity, generative systems
can refine stroke sequencing, rhythmic structure, and expressive
parameters directly within the CWSR representation. This opens a path
toward computational models that internalize calligraphic knowledge as
performative rules rather than implicit visual patterns.
 
 \subsection{Applications}

CWSR enables a range of practical and research applications:

\begin{enumerate}
\item \textbf{Computational Analysis:} quantitative investigation of structural balance, stroke connectivity, rhythmic pacing, and temporal flow in calligraphic composition.
\item \textbf{Generative Systems:} controllable synthesis of calligraphic writing through interpretable structural representations and explicit style parameters.
\item \textbf{Pedagogy:} teaching calligraphy as a structured and embodied practice, emphasizing rhythm, stroke sequencing, and compositional organization.
\item \textbf{Digital Preservation:} capturing dynamic and performative aspects of calligraphy that are not preserved in static images alone.
\item \textbf{Performative Media:} real-time visualization, sonification, or robotic execution of calligraphy scores in interactive or artistic environments.
\end{enumerate}
 
\subsection{Future Directions}

Several directions may further extend the CWSR framework:

\begin{itemize}
\item Integration with modern generative models as a controllable structural or conditioning layer.
\item Development of score-based similarity metrics that compare calligraphic works based on execution structure rather than visual resemblance alone.
\item Extension of the framework to other embodied writing and drawing practices, including cursive handwriting, sketching, and ritualized inscription.
\item Exploration of real-time interactive systems for teaching, digital preservation, and live artistic performance.
\item Further development  of computational agents capable of generating executable CWSR scores, enabling systematic exploration of stylistic variation and performative strategies.
\end{itemize}
  
\section{Conclusion}

This work introduces \textbf{Calligraphy Writing Score Representation (CWSR)} and
articulates \textbf{Shu Dao} as a performance-centered conceptual framework for
East Asian calligraphy. Rather than viewing calligraphic characters as static
visual artifacts, Shu Dao reframes writing as an embodied and time-based
practice in which meaning emerges through ordered action, rhythm, and
expressive control.

CWSR operationalizes this perspective by encoding each stroke as a performable
unit within a symbolic score. Stroke type, spatial placement, execution order,
trajectory, and dynamic attributes such as pressure and pacing are explicitly
represented, enabling both humans and computational systems to interpret,
execute, and regenerate calligraphy from the same structured description. In
this way, structural identity is separated from expressive realization,
analogous to the relationship between musical notation and musical
performance.

By making temporal structure and performer intention explicit, CWSR exposes
dimensions of calligraphic knowledge that are traditionally transmitted
implicitly through practice and apprenticeship. The framework enables
interpretable computational analysis, controllable generative modeling,
pedagogical visualization, and digital preservation that extend beyond
conventional image-based representations.
 
 More broadly, this work positions Chinese calligraphy and related
East Asian traditions alongside music and ritual performance as
score-mediated art forms. Through Shu Dao and its implementation
in CWSR, calligraphy is presented not only as visual composition
but also as a structured, performable, and generative practice.
This perspective provides a foundation for future research in
computational analysis of writing gestures, generative calligraphy
systems, and AI-assisted calligraphy.

\section {Declaration of generative AI and AI-assisted technologies in the manuscript preparation process}
Statement: During the preparation of this work the authors used ChatGPT   in order to prepare and writing the draft paper. After using this tool, the authors reviewed and edited the content as needed and take full responsibility for the content of the published article.

\bibliographystyle{plain}
 
\bibliography{shudao}

  \appendix
\section{Example Calligraphy Writing Score Representation (CWSR)}
\label{app:cwsr-example}
 This appendix presents a complete example of a
\emph{Calligraphy Writing Score Representation (CWSR)} in both
machine-readable and human-readable forms. The score encodes not only
stroke geometry, but also rhythm, gesture, material properties, and
performance-related attributes. It is designed to function analogously
to a musical score or a \textit{Chadao} (tea ceremony) procedural script:
a structured specification that enables interpretation, reenactment,
and computational generation.

The example corresponds to the phrase ``永和九年'', rendered in a style
inspired by Wang Xizhi. It demonstrates how static calligraphic forms
can be extended into a temporally grounded representation that captures
the performative dynamics of calligraphic writing.

\subsection*{A.1 Machine-Readable Score (JSON Representation)}
This JSON-based representation provides a precise and unambiguous
encoding of the CWSR score in a machine-readable format.

\includepdf[
    pages=-,
    scale=0.9,
    pagecommand={}
]{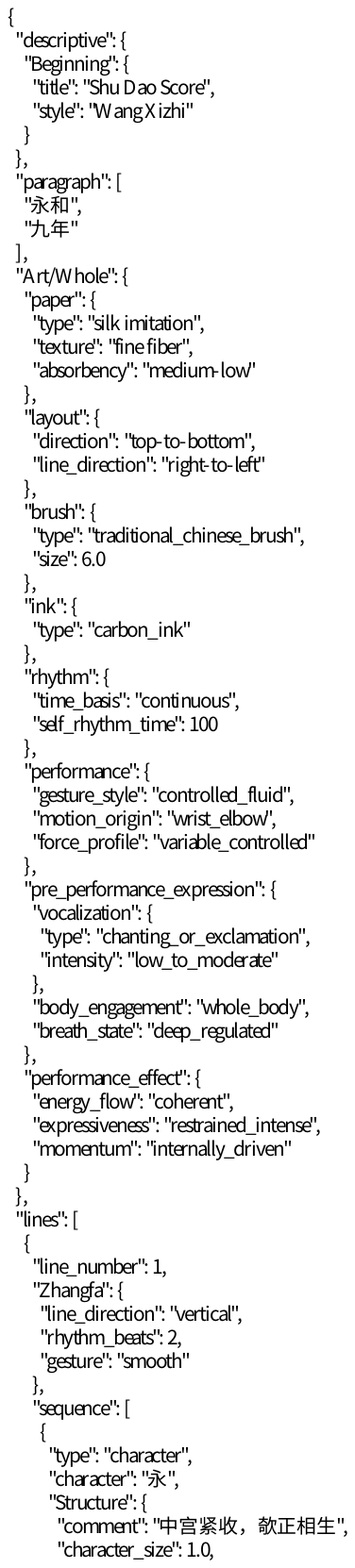}

 \subsection*{A.2 Human-Readable Symbolic Score}

This section presents the same CWSR score shown in Appendix~A.1,
but in a human-readable symbolic notation. While Appendix~A.1 provides
a machine-readable JSON representation for computational processing,
the symbolic score is designed for direct human reading,
interpretation, and performance.

Like a musical score, it exposes stroke sequence, spatial structure,
and dynamic emphasis in a visually interpretable form, allowing
practitioners to understand and perform the writing without requiring
computational execution.

\includepdf[
    pages=-,
    scale=0.9,
    pagecommand={}
]{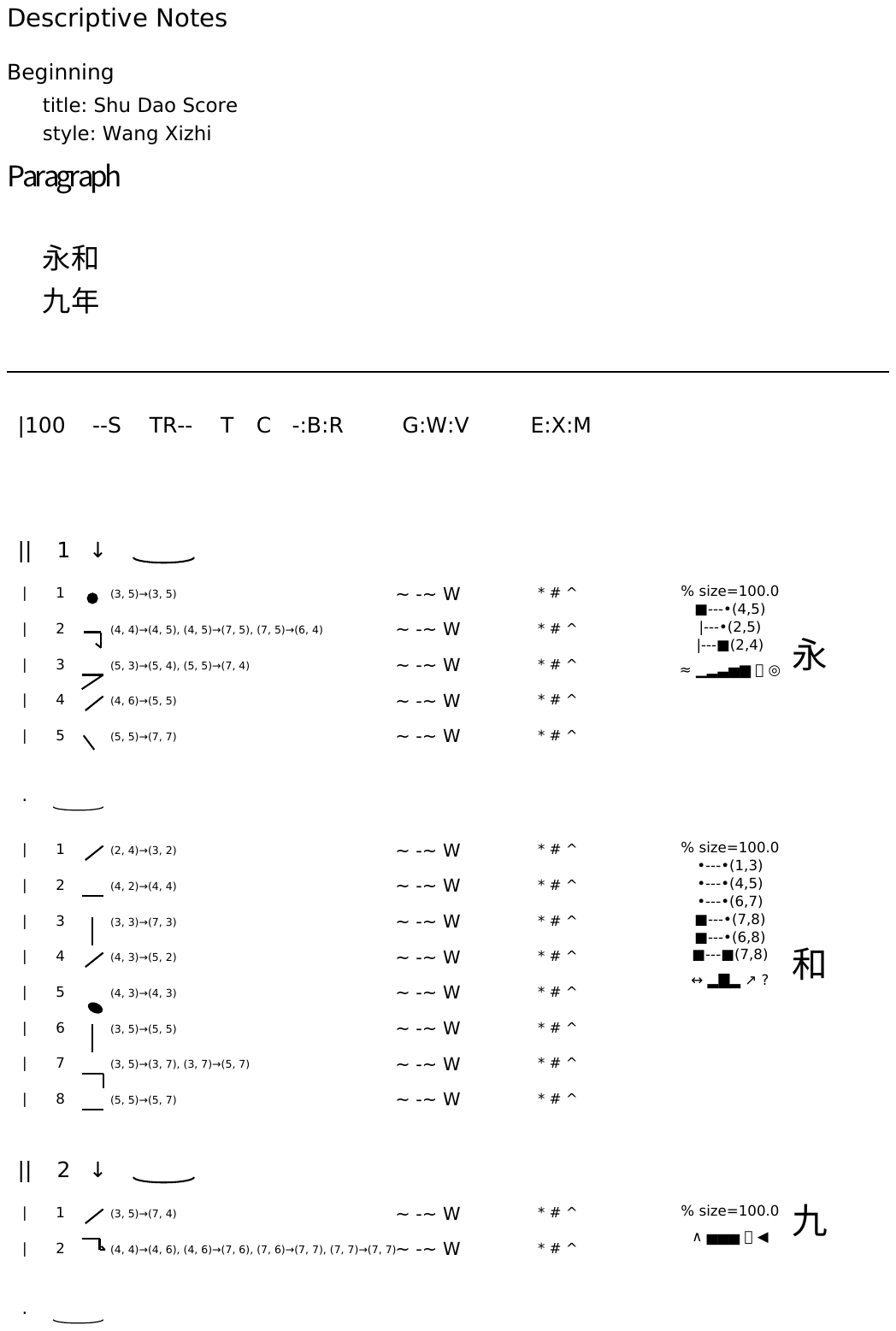}

  \subsection*{A.3 AI-Agent-Generated Score}

This section presents an executable calligraphy score in JSON format
generated by an AI agent based on the CWSR score shown in
Appendix~A.1. The agent produces a complete Score JSON that encodes
the writing process as a sequence of temporally ordered actions,
including stroke types, relative timing, directional flow, pressure
profiles, and gesture-level constraints.

The resulting representation is directly executable by a rendering
or simulation system, allowing the score to be interpreted,
visualized, or performed without reference to source images.

 \includepdf[
    pages=-,
    scale=0.9,
    pagecommand={}
]{Wang_xizhi_shu_dao_ai_score.pdf}

\subsection*{A.4 Character Rendering from Executable Score JSON}

The following Python function illustrates how an executable Score JSON,
produced by an AI agent, is rendered into SVG calligraphy. The rendering
uses both the structural information (strokes, connections) and AI-provided
stylistic hints (motion, pressure, gesture, emphasis, and ink), which are
automatically filled if missing.

For brevity, we show the implementation for only one stroke type (\texttt{heng}).
Other stroke types (\texttt{na}, \texttt{dian}, etc.) follow a similar approach.

\begin{lstlisting}[language=Python, caption={Drawing a character using AI-enhanced features from the executable Score JSON.}, label={lst:draw_character}]
def draw_character(svg, x, y, size, item):
    """
    Draw a character in SVG at (x, y) with given size.
    
    Uses strokes and optional AI-provided 'hint' features
    (motion, pressure, gesture, emphasis, ink), automatically filled if missing.
    """
    g = SubElement(svg, "g", transform=f"translate({x},{y})")

    # Optional character bounding box for debugging
    SubElement(
        g,
        "rect",
        x="0",
        y="0",
        width=str(size),
        height=str(size),
        fill="none",
        stroke="#ccc"
    )

    structure = item.get("Structure", {})
    connections = structure.get("connections", {})

    # Extract AI hints or use defaults
    hint = structure.get("hint", {})
    motion = hint.get("motion", "flowing")
    pressure = hint.get("pressure", "medium")
    gesture = hint.get("gesture", "press-lift")
    emphasis = hint.get("emphasis", "central")
    ink = hint.get("ink", {"wetness": 0.8, "feibai": 0.1})

    for stroke in item.get("strokes", []):
        # Merge stroke-level hints if present; fallback to character-level
        width_variation = stroke.get("width_variation", {"start": 2, "peak": 2.4, "end": 0.8})
        stroke_gesture = stroke.get("gesture_trajectory", gesture)
        stroke_ink = stroke.get("ink", ink)

        stroke_width = stroke.get("stroke_width", 2.0)
        stroke_pressure = stroke.get("pressure_curve", PRESSURE_MAP.get(pressure, PRESSURE_MAP["medium"]))

        # Draw stroke with all parameters
        draw_stroke(
            g,
            stroke,
            connections,
            size,
            motion=motion,
            pressure_curve=stroke_pressure,
            width_variation=width_variation,
            gesture=stroke_gesture,
            emphasis=emphasis,
            stroke_width=stroke_width,
            ink=stroke_ink
        )
\end{lstlisting}

 \begin{lstlisting}[language=Python, caption={Rendering ink samples with width and pressure interpolation in SVG}, label={lst:render_ink}]
def render_ink(svg_group, samples, params=None, default_opacity=0.8, default_color="black"):
    """
    Render gesture samples as ink dots in SVG using width/pressure/ink from params.
    """
    # fallback width if no params provided
    ws = {"start": 2.0, "peak": 2.5, "end": 1.0}  
    pressures = {"start": 0.6, "peak": 0.8, "end": 0.3}

    if params is not None:
        ws = {
            "start": params.get("width_start", ws["start"]),
            "peak": params.get("width_peak", ws["peak"]),
            "end": params.get("width_end", ws["end"])
        }
        pressures = {
            "start": params.get("pressure_start", pressures["start"]),
            "peak": params.get("pressure_peak", pressures["peak"]),
            "end": params.get("pressure_end", pressures["end"])
        }

    n = len(samples)
    for i, s in enumerate(samples):
        t = i / max(1, n - 1)  # normalized 0~1 along stroke
        # simple quadratic interpolation: start → peak → end
        if t < 0.5:
            width = ws["start"] + (ws["peak"] - ws["start"]) * (t / 0.5)
            opacity = pressures["start"] + (pressures["peak"] - pressures["start"]) * (t / 0.5)
        else:
            width = ws["peak"] + (ws["end"] - ws["peak"]) * ((t - 0.5) / 0.5)
            opacity = pressures["peak"] + (pressures["end"] - pressures["peak"]) * ((t - 0.5) / 0.5)

        SubElement(
            svg_group,
            "circle",
            cx=f"{s['x']:.2f}",
            cy=f"{s['y']:.2f}",
            r=f"{width / 2:.2f}",       # radius = half width
            fill=default_color,
            opacity=f"{opacity:.3f}"
        )
\end{lstlisting}

\begin{lstlisting}[language=Python, caption={Rendering a horizontal stroke (`heng`) in SVG using stroke geometry and AI-generated hints}, label={lst:draw_heng}]
def draw_heng(
    svg_group,
    stroke,
    connections,
    size,
    **kwargs
):
    """
    横 (heng):
    - slight upward lift in the middle
    - geometry only
    - pressure / width / ink driven by kwargs
    """
    params = normalize_gesture_kwargs(kwargs)
    stroke_width = params.get("stroke_width", 2.0)  # fallback default if missing
    sx, sy = map_point(stroke["start"], size)
    ex, ey = map_point(stroke["end"], size)
    
   
    #   微提中锋（横画呼吸感）
       
    lift = size * 0.03

    points = [
        (sx, sy),                    # 起笔
        ((sx + ex) / 2, sy - lift),  # 行笔微提
        (ex, ey)                     # 收笔
    ]
 
    # ---- gesture + ink simulation ----
    samples = simulate_gesture(points, params)

    # ---- render ----
    render_ink(svg_group, samples, params=params)
\end{lstlisting} 
  
\end{document}